\documentclass[nobibnotes,nofootinbib]{revtex4}
\usepackage{epsfig}
\usepackage{amsmath,amssymb,bm}
\newcommand{\calN}{{\cal N}}
\newcommand{\rr}{\bm{r}}
\newcommand{\bb}{\bm{b}}
\newcommand{\xx}{\bm{x}}
\newcommand{\yy}{\bm{y}}

\def\beq{\begin{equation}}
\def\eeq{\end{equation}}
\begin{document}
\title{Heavy Quark Production in Ultra High Energy Cosmic Ray Interactions}
\author{V.P. Gon\c{c}alves$^{1}$\footnote{Corresponding author. Email: barros@ufpel.edu.br} and D. R. Gratieri$^{1,2}$}
\affiliation{$^{1}$
Instituto de F\'{\i}sica e Matem\'atica,  Universidade
Federal de Pelotas, 
Caixa Postal 354, CEP 96010-900, Pelotas, RS, Brazil}

\affiliation{$^{2}$
Instituto de F\'isica Gleb Wataghin - UNICAMP, 13083-859, Campinas, SP, Brazil\ 
}

\begin{abstract}
In this paper we present a comprehensive study of the heavy quark production in ultra high energy cosmic ray interactions in the atmosphere considering that the primary cosmic ray can be either a photon, neutrino or a proton. The analysis is performed using a unified  framework -- the dipole formalism --- and  the  saturation effects, associated to the physical process of parton recombination, are taken into account. We demonstrate that the contribution of heavy quarks for cosmic ray interactions is in general non-negligible and can be dominant depending of the process considered. Moreover, our results indicate that new dynamical mechanisms  should be included in order to obtain reliable predictions  for the heavy quark production in $pp$ collisions at ultra high cosmic ray energies. 
 
\end{abstract}
%
%
\maketitle
%
%
%

\section{Introduction}
  Ultra high energy cosmic rays (UHECRs) remains
a puzzle in physics. Although the existence of UHECRs with energies
above $10^{11}$ GeV is now a well-established fact, the theoretical
understanding of its origin and propagation is a subject of  strong
interest and intense discussion \cite{review_cosmic}.  In particular, the recent detection of  ultra-high energies (UHE) neutrinos by the IceCube Neutrino Observatory \cite{Icecube} starts a new era in the neutrino physics. These and forthcoming data from IceCube and Pierre Auger Observatories may shed light on the Standard Model as well as  reveal aspects of new physics \cite{review_neutrinos}.  
An important subject in cosmic ray physics is the flux of prompt
leptons at the Earth which is related to   primary interactions at
energies that can by far exceed the highest available in accelerators. This flux is directly associated with the charmed particle production and its 
decays, being  its estimation strongly dependent on the model used to
calculate the charm production cross section and energy spectra
\cite{leptonflux}.  The quantification of the prompt leptons fluxes
is essential, for instance, for neutrino physics, since the flux of
prompt neutrinos provides the main background of the muon neutrinos
from extra-galactic neutrino sources, which are being studied by the  IceCube Observatory.


In this paper we calculate the heavy quark production in UHECR interactions considering that the particle incident in the atmosphere can be either a photon, neutrino or a proton. As this interaction probes the theory of the strong interactions in a
new kinematical range characterized by a center of mass energy of
approximately 500 TeV, which is more than one order of magnitude
larger than that of the Large Hadron Collider at CERN, we take into account the nonlinear corrections to the QCD dynamics effects associated to the gluon saturation physics \cite{book_kov}. In this energy regime, 
 perturbative Quantum Chromodynamics (pQCD) predicts that the small-$x$ gluons in a hadron wavefunction should form a Color Glass Condensate (CGC),  which  is characterized by the limitation on the maximum
phase-space parton density that can be reached in the hadron
wavefunction (parton saturation), with the transition being
specified  by a typical scale, which is energy dependent and is
called saturation scale $Q_{\mathrm{sat}}$. Moreover, we calculate the heavy quark production using the
color dipole approach, which gives a simple unified picture for this
process in photon-proton, neutrino - proton and proton - proton interactions. 
Our goal is to  extend the  analysis performed in Ref. \cite{victor_jhep} some years ago, where the main focus was to estimate the heavy quark cross section  in photon - air and proton - air interactions   taking into account nuclear effects.
In particular, we extend the study for  neutrino - proton interactions and explicitly estimate, for the first time,  the  heavy quark contributions for the total $\gamma p$, $\nu p$ and $pp$ cross sections. 
Distinctly from Ref.  \cite{victor_jhep}, the $\gamma p$ and $pp$ interactions are analysed in detail, with the free parameters present in the color dipole predictions being constrained by the experimental data for the total $\gamma p$ cross section and by the recent LHC data for the heavy quark production, which strongly reduces the uncertainty in  the predictions for ultra high cosmic ray energies. Moreover, the  heavy quark production in a double parton scattering process in cosmic ray $pp$ interactions is estimated for the first time.

 It is important to emphasize that although there are another studies in the literature that calculate the heavy quark contribution for the total cross sections (See {\it e.g.} Refs. \cite{kms,reno,frankfurt,Marta_Rafal}), it is the first time that it is estimated using a unified approach to treat the distinct processes. In general, these studies have been performed for a specific process of the three  analysed here  considering different underlying assumptions, as {\it e.g.} the order of the perturbative calculations,  the QCD evolution equations and/or choice of the parton distributions, heavy quark masses and factorization scale, which introduces a large theoretical uncertainty in the predictions. Distinctly of these previous studies, our predictions for the heavy quark contribution in $\gamma p$, $\nu p$ and $pp$ processes are based on a unified approach, considering a unique model for the QCD dynamics and for the value of the heavy quark masses.

The paper is organized as follows. In the next Section we present a brief review of the  color dipole picture, presenting the main formulas and assumptions present in the calculation of the heavy quark production in this approach. In Section \ref{dynamics_section} we discuss the QCD dynamics at high energies and present the models used for the dipole - proton scattering amplitude used in our calculations. In Section \ref{results_section} we present our results for the energy dependence of the heavy quark contributions for the  total  $\gamma p$, $\nu p$ and $pp$ cross sections. A comparison with previous predictions is presented and the magnitude of the gluon saturation effects is estimated. Moreover, the probability of the heavy quark production in a  double parton scattering process at ultra high cosmic ray $pp$ interactions is calculated.  Finally, in Section \ref{summary_section} we summarize  our main conclusions.

\section{Heavy quark production in the dipole picture}
\label{hq_section}

 Let us start our
analysis considering the  photon-hadron
interactions at high energies. It  is usually described in the
infinite momentum frame  of the hadron in terms of the scattering of
the photon off a sea quark, which is typically emitted  by the
small-$x$ gluons in the proton. However, in order to disentangle the
small-$x$ dynamics of the hadron wavefunction, it is more adequate
to consider the photon-hadron scattering in the dipole frame, in
which most of the energy is carried by the hadron, while the  photon
has just enough energy to dissociate into a quark-antiquark pair
before the scattering. In this representation the probing projectile
fluctuates into a quark-antiquark pair (a dipole) with transverse
separation $\rr$ long after the interaction, which then scatters off
the proton \cite{nik}. In this approach  the heavy quark
photoproduction cross section  reads as,
\begin{eqnarray}
\sigma^f_{\gamma p}(W^2)=  \int dz\, d^2\rr
\,|\Psi^\gamma_{f,T} (z,\,\rr)|^ 2 \,\sigma_{dip}(\rr, x) ,
\label{dipapprox}
\end{eqnarray}
where $W^2$ is the squared photon - hadron center-of-mass energy and  $\Psi^{\gamma}_{f,T}(z,\,\rr)$  is the transverse light-cone
wavefunction of the photon for a heavy quark dipole of flavor $f$  \cite{nik_neu}.
Moreover,  $z$ and $\bar{z}=(1-z)$ are the longitudinal momentum fractions of the quark and antiquark, respectively. The dipole cross section, $\sigma_{dip}$, parametrizes the cross section of interaction dipole-proton. As usual, the Bjorken variable is denoted by $x$.

In the case of deep inelastic neutrino - proton scattering, it  is described in terms of charged current (CC) and neutral current (NC) interactions, which proceed through $W^{\pm}$ and $Z^0$  exchanges, respectively.   The total neutrino - proton cross section is given by \cite{book}
\begin{eqnarray}
\sigma_{\nu p} (E_\nu) = \int_{Q^2_{min}}^s dQ^2 \int_{Q^2/s}^{1} dx \frac{1}{x s} 
\frac{\partial^2 \sigma_{\nu p}}{\partial x \partial y}\,\,,
\label{total}
\end{eqnarray}
where $E_{\nu}$ is the neutrino energy, $s = 2 ME_{\nu}$ with $M$ the proton mass, $y = Q^2/(xs)$ and $Q^2_{min}$ is the minimum value of $Q^2$ which is introduced in order to stay in the deep inelastic region. In what follows we assume $Q^2_{min} = 1$ GeV$^2$. Moreover, the differential cross section is given by \cite{book}
\begin{widetext}
\begin{eqnarray} 
\frac{\partial^2 \sigma_{\nu p}}{\partial x \partial y} = \frac{G_F^2 M E_{\nu}}{\pi} \left(\frac{M_i^2}{M_i^2 + Q^2}\right)^2 \left[\frac{1+(1-y)^2}{2} \, F_{2}^\nu (x,Q^2) - \frac{y^2}{2}F_{L}^\nu(x,Q^2)+ y (1-\frac{y}{2})xF_{3}^\nu(x,Q^2)\right]\,\,,
\label{difcross}
\end{eqnarray}
\end{widetext}
where  $G_F$ is the Fermi constant and the mass $M_i$ is either $M_W$ or $M_Z$ according to whether we are calculating CC or NC neutrino interactions. 
The calculation of $\sigma_{\nu p}$ involves  integrations over $x$ and $Q^2$, with the integral being dominated by  the interaction with partons of lower $x$ and  $Q^2$ values of the order of the electroweak boson mass squared.
In the color dipole picture  the  $F_2^\nu$ structure function is expressed in terms of the transverse and longitudinal structure functions, $F_2^\nu=F_T^\nu + F_L^\nu$, which are given by 
 \begin{widetext}
\begin{eqnarray}
&\,& F_{T,L}^\nu (x,Q^2)  = \frac{Q^2}{4\pi^2} \sum_f  \int_0^1 dz \int d^2  \rr |\Psi^{W,Z}_{f,T,L}(\rr,z,Q^2)|^2 \sigma_{dip}(\rr,x)\,\,
\label{funcs}
\end{eqnarray} 
\end{widetext}
where the functions  $\Psi^{W,Z}_{T,L}$ are  the wave functions of the virtual gauge bosons corresponding to their transverse or longitudinal polarizations \cite{nik_neu}. For NC interactions, the neutral gauge boson can fluctuate in a dipole $f\bar{f}$ with a given flavour $f$ $(= u, d, s, c, b, t)$. On the other hand, for CC interactions  we assume that the charged gauge boson can fluctuate in a dipole $f_i\bar{f}_j$ ($i \neq j$) with one of the following combinations of flavours: $ud$, $cs$ or $bt$.

Concerning heavy quark hadroproduction it  can also be described
in terms of the color dipole cross section, similarly to
photon/neutrino-hadron interactions, using the color dipole approach
\cite{npz}. In this approach the total heavy quark production cross
section is given by \cite{npz}
\begin{eqnarray}
\sigma(pp\to Q\bar Q X) & = &
2\int_0^{-\ln(\frac{2m_Q}{\sqrt{s}})}dy\,
x_1G\left(x_1,\mu_F^2\right) \times \nonumber \\
& \times &  \sigma(GN\to Q\bar Q X)\,\,,
\label{ccppdip}
\end{eqnarray}
where  $y$ is the rapidity of the pair,
$\mu_F\sim m_Q$ is the factorization scale, $x_1G(x_1,\mu_F^2)$ is the projectile gluon density at scale $\mu_F$ and longitudinal momentum fraction $x_1$ and the
partonic cross section $\sigma(GN\to Q\bar Q X)$ is given by
\beq\label{eq:all} \sigma(GN\to Q\bar Q X) =\int dz \, d^2 \rr
\left|\Psi_{G\to Q\bar Q}(z,\rr)\right|^2 \sigma_{q\bar q G}(z,\rr),
\nonumber \eeq with $\Psi_{G\to Q\bar Q}$ being the pQCD calculated
distribution amplitude, which describes the dependence of the $|Q
\bar Q \rangle$ Fock component on transverse separation and
fractional momentum. Moreover, $\sigma_{q\bar qG}$ is the cross
section for scattering a color neutral quark-antiquark-gluon system
on the proton and is directly related with the dipole cross section
as follows
 \beq\label{eq:qqG}
\sigma_{q\bar qG}=\frac{9}{8}\left[\sigma_{dip}(x_2,z \rr)
+\sigma_{dip}(x_2,\bar{z}\rr)\right] -\frac{1}{8}\sigma_{dip}(x_2,\rr)
\nonumber. \eeq  The basic idea on this approach is that at high
energies a gluon $G$ from the projectile hadron can develop a
fluctuation which contains a heavy quark pair ($Q\bar Q$).
Interaction with the color field of the target then may release
these heavy quarks. 

A comment is order here. In contrast to the standard calculations of the heavy quark contributions for the total cross sections, which are based on the  collinear factorization at leading twist and the solution of the linear DGLAP equations, the cross sections obtained in the color dipole approach  resums higher-twist  corrections beyond the traditional factorization schemes and allow us to include the gluon saturation effects in a straightforward and unified way. However, it is important to emphasize that still is not clear if the dipole factorization should not be generalized at  larger than the current collider energies. Results obtained in Ref. \cite{raju} for the quark pair or gluon production in the interaction of two ultra dense systems indicate the breakdown of the factorization at high energies. As the proton gluon density for ultra high cosmic ray energies is expected to be very large, a similar scenario can be present in $pp$ cosmic ray interactions. However, if and for what energy this breakdown occurs still are open questions which deserve more detailed studies.

 \section{High
Energy QCD Dynamics}
\label{dynamics_section}
 
As demonstrated in the previous section, in the color dipole formalism the  heavy quark cross sections are determined by the dipole - proton cross section, $\sigma_{dip}$, which 
 encodes all the information about the hadronic scattering, and thus about the nonlinear and quantum effects in the hadron wave function. It can be expressed by
\begin{equation}\label{eq:dipcross}
\sigma_{dip}(\rr,x)=2\int d^2b\,\calN(\bb,\rr,x),
\end{equation}
where $\calN(\bb,\rr,x)$ is the imaginary part of the forward amplitude for the scattering between a small dipole
(a colorless quark-antiquark pair) and a dense hadron target, at a given
rapidity interval $Y=\ln(1/x)$. The dipole has transverse size given by the vector
$\rr=\xx-\yy$, where $\xx$ and $\yy$ are the transverse vectors for the quark
and antiquark, respectively, and impact parameter $\bb=(\xx+\yy)/2$. 
At high energies the evolution with the rapidity $Y$ of
$\calN(\rr,\bb,Y)$  is given by the infinite hierarchy of equations, the so called
Balitsky-JIMWLK equations \cite{bal,cgc}, which reduces in the mean field approximation to the Balitsky-Kovchegov (BK) equation \cite{bal,kovchegov}. It is useful to assume  the translational invariance approximation, which regards hadron homogeneity in the
transverse plane. It implies that the dipole-proton cross section reads $\sigma_{dip}(\rr,x)=\sigma_0\calN(\rr,x)$, 
where the constant $\sigma_0$, which results from the $\bb$ integration, sets the normalization.
Moreover,  the amplitude becomes independent of the impact parameter
$\bb$ and depends only on the dipole size $r=|\rr|$, i.e. $\calN_Y(\rr)=\calN_Y(r)$.
Although a complete analytical solution of the BK equation is still lacking, its main properties are known:
(a) for the interaction of a small dipole ($\rr \ll
1/Q_{\mathrm{sat}}$), ${\cal{N}}(\rr) \approx \rr^2$, implying  that
this system is weakly interacting; (b) for a large dipole ($\rr \gg
1/Q_{\mathrm{sat}}$), the system is strongly absorbed and therefore
${\cal{N}}(\rr) \approx 1$. The typical momentum scale, $Q_{\mathrm{sat}}^2\propto x^{-\lambda}\,(\lambda\approx 0.3)$, is the so called saturation scale. This property is associated  to the
large density of saturated gluons in the hadron wave function. 
  In our analysis we will consider the  phenomenological saturation model
proposed in Ref. \cite{GBW}, denoted GBW hereafter, which encodes the main properties of the
saturation approaches, with the scattering amplitude  parametrized
as follows
\begin{equation}\label{eq:gbw}
\calN(r,Y)=1-e^{-r^2Q_{\mathrm{sat}}^2(Y)/4},
\end{equation}
where the saturation scale is given by $Q_{\mathrm{sat}}^2=Q_0^2\left(x_0/x\right)^{\lambda}$,
$x_0$ is the value of the Bjorken $x$ in the beginning of the evolution and $\lambda$ is the
saturation exponent. The parameters $\sigma_0$, $x_0$ and $\lambda$ are obtained by fitting the $ep$ HERA data. In our study we assume the values obtained in Ref. \cite{Koslov07}, where the GBW model was updated in order to describe more recent data.  It is important to emphasize that we have verified that our main conclusions are not modified if another phenomenological saturation model or the solution of the BK equation, as given  in Ref. \cite{rcbk}, are used as input in our calculations.

\begin{figure}[t]
\vspace{0.5cm}
\includegraphics[ scale=0.25]{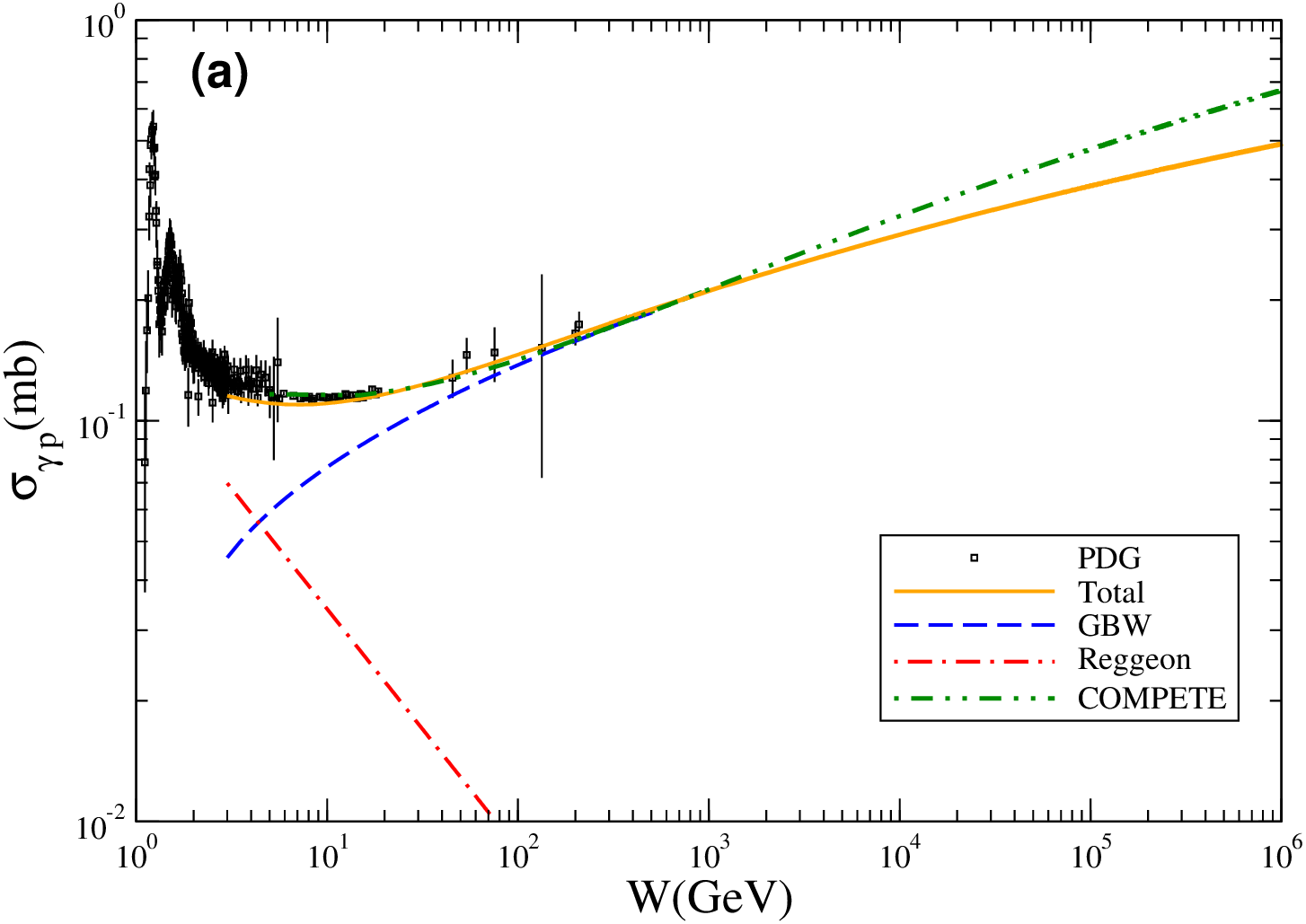}
\includegraphics[scale=0.25]{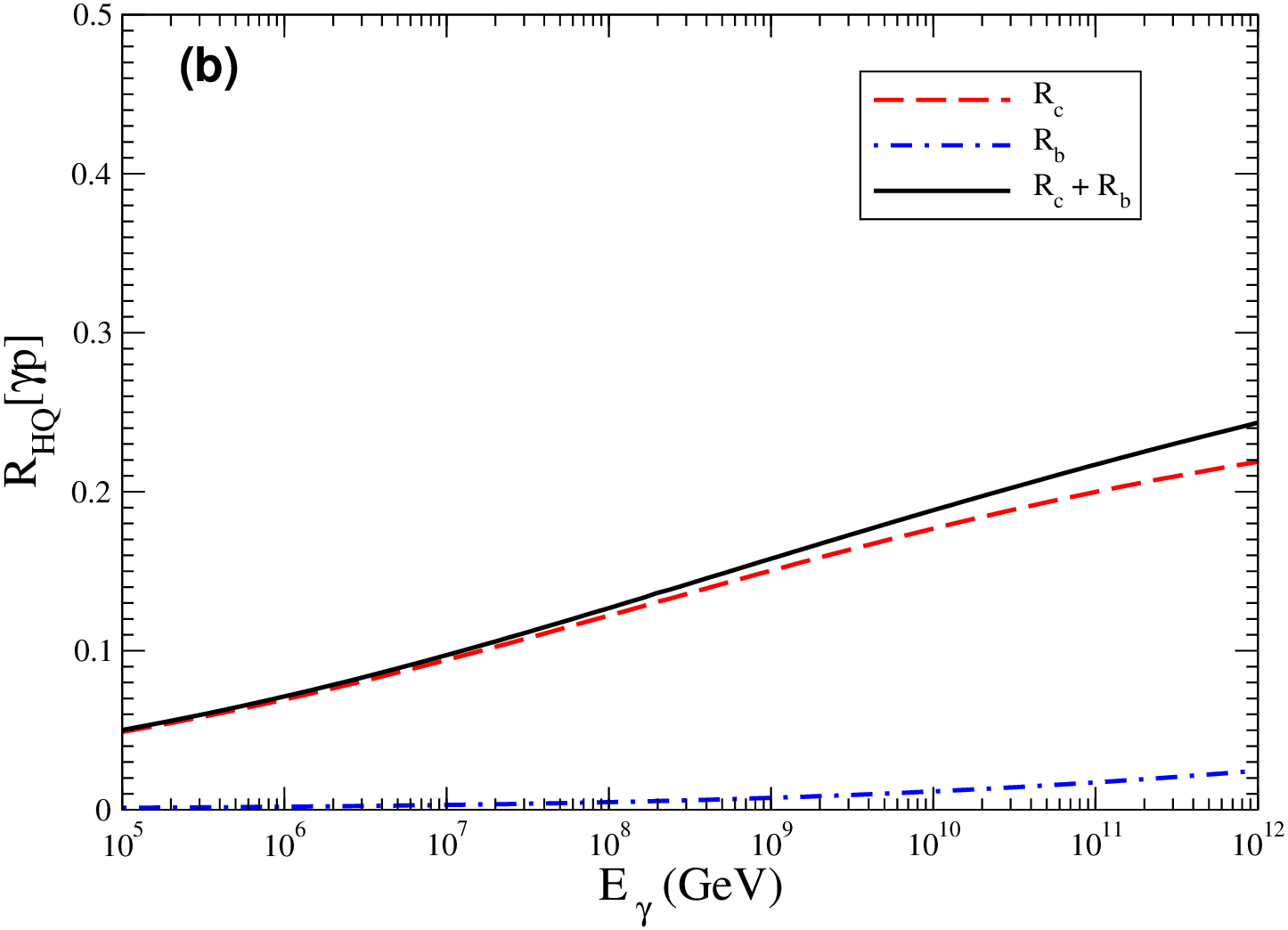}
\caption{ (Color online) (a) The total $\gamma p $ cross section as a function of the center-of-mass energy. Data from Ref. \cite{pdg}. (b) The ratio $R_{HQ}[\gamma p]$ as a function of the energy $E_{\gamma}$ of the primary photon.  }
\label{fig:gamap}
\end{figure}


\begin{figure}[t]
\vspace{0.5cm}
\includegraphics[scale=0.25]{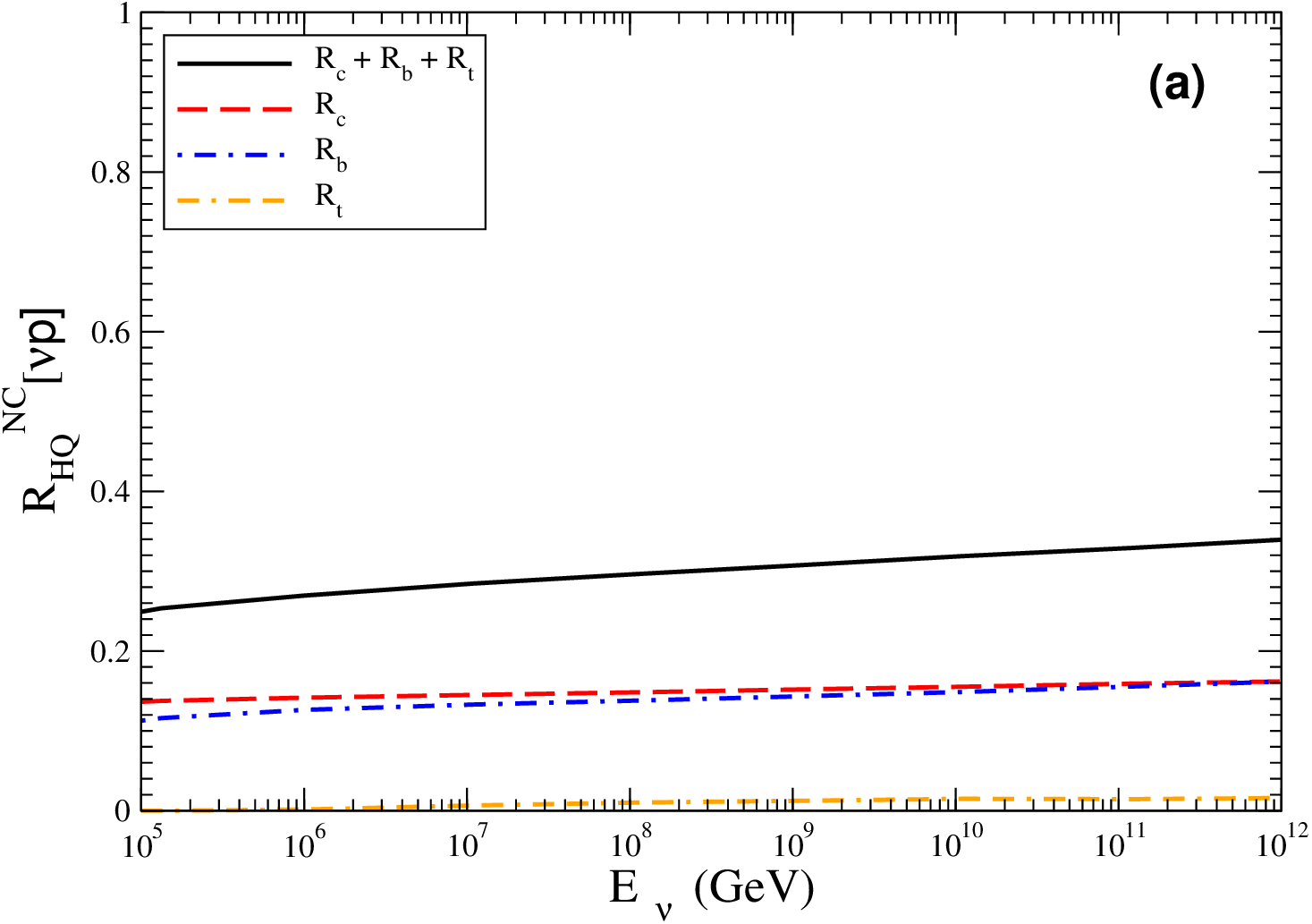} 
 \includegraphics[scale=0.25]{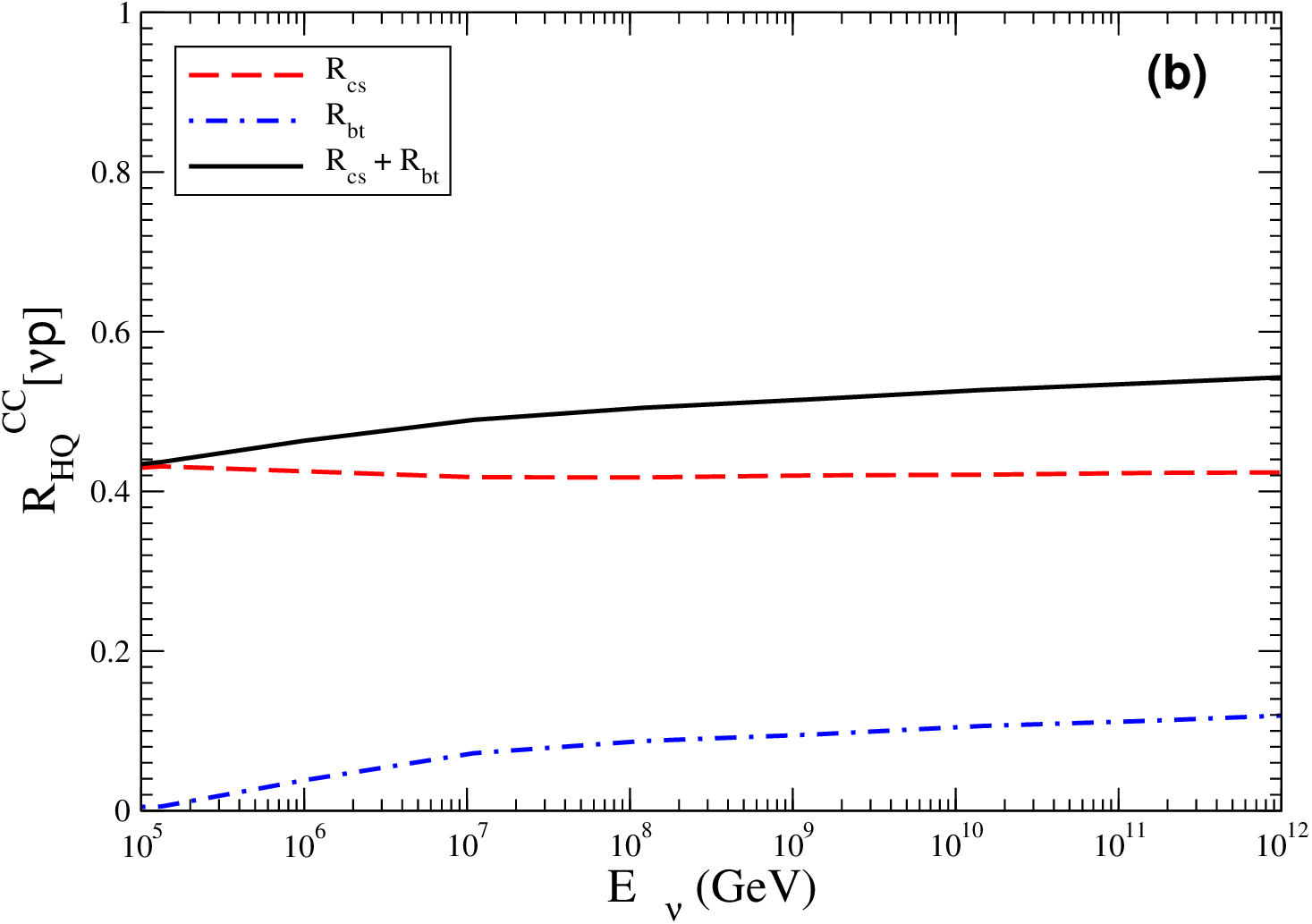}
\caption{ (Color online) (a) The ratio $R_{HQ}[\nu p]$ as a function of the energy $E_{\nu}$ of the primary neutrino for NC interactions. (b) The ratio $R_{HQ}[\nu p]$ as a function of the energy $E_{\nu}$ of the primary neutrino for CC interactions. }
\label{fig:neutrinop}
\end{figure}

\begin{figure}[t]
\vspace{0.5cm}
\includegraphics[scale=0.25]{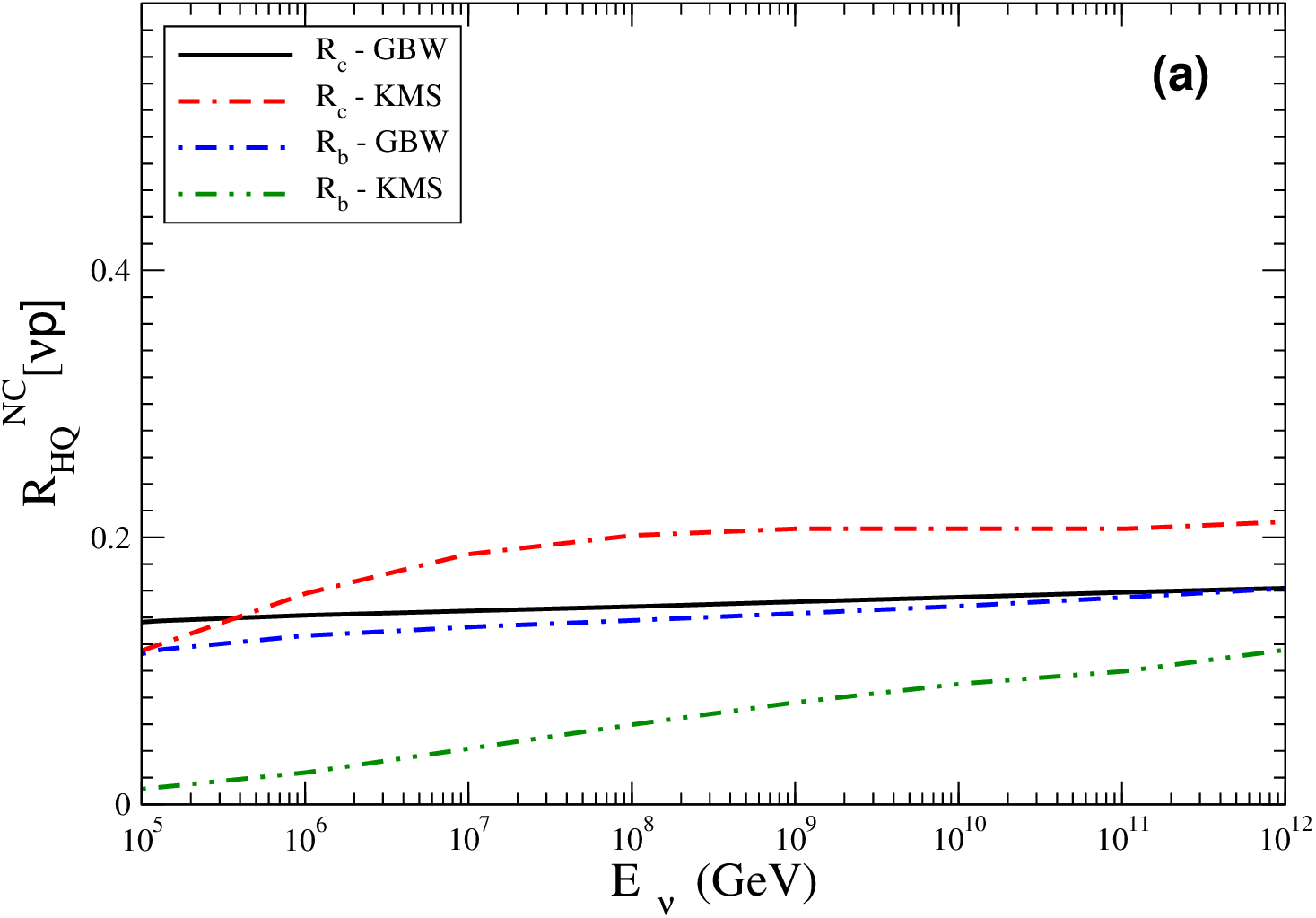} 
 \includegraphics[scale=0.25]{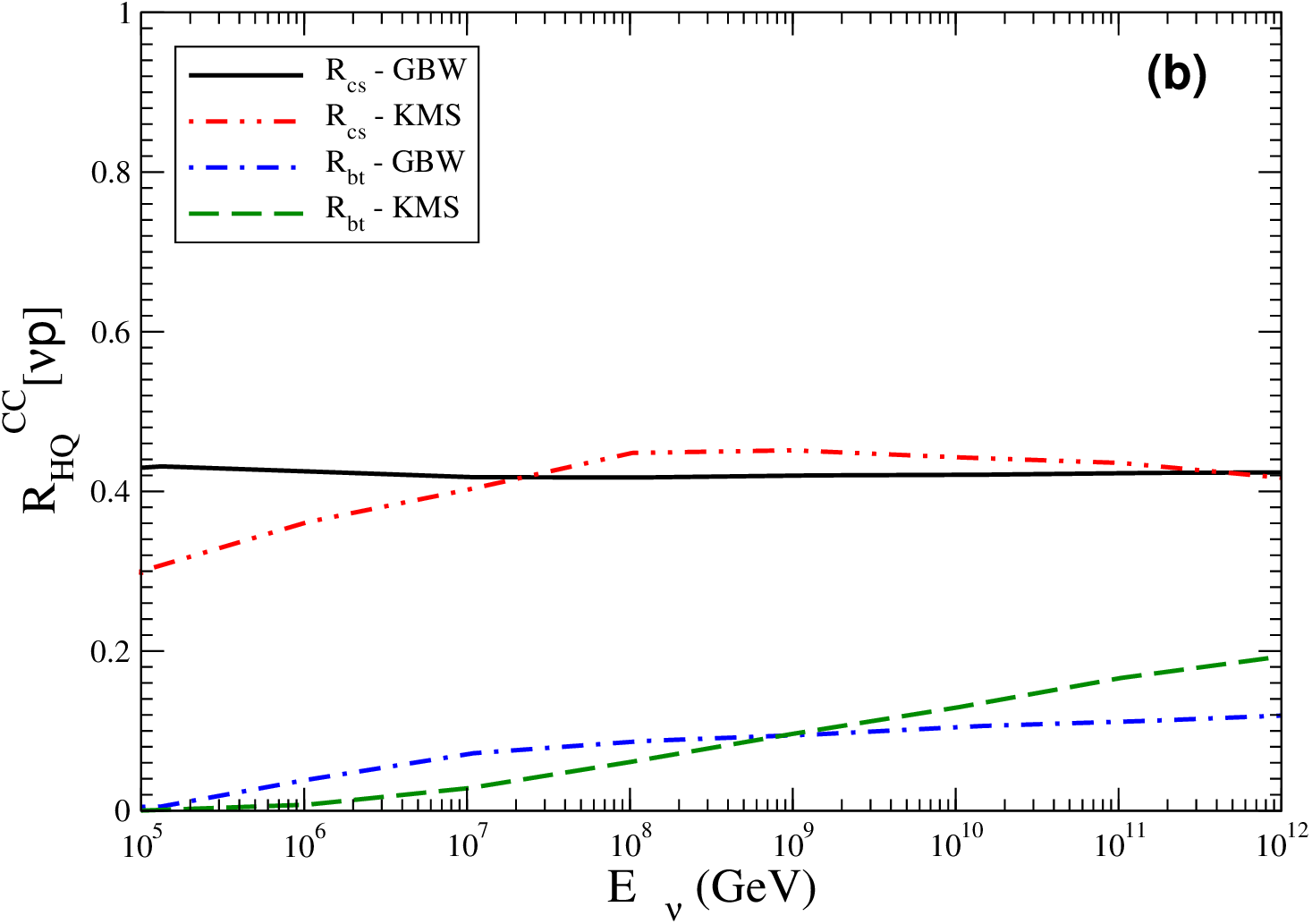}
\caption{ (Color online) Comparison between the GBW and KMS predictions for the ratios  (a) $R_{HQ}^{NC}[\nu p]$ and  (b) $R_{HQ}^{CC}[\nu p]$ as a function of the energy $E_{\nu}$ of the primary neutrino. }
\label{fig:neutrinop2}
\end{figure}

\begin{figure}[t]
\vspace{0.5cm}
\includegraphics[scale=0.25] {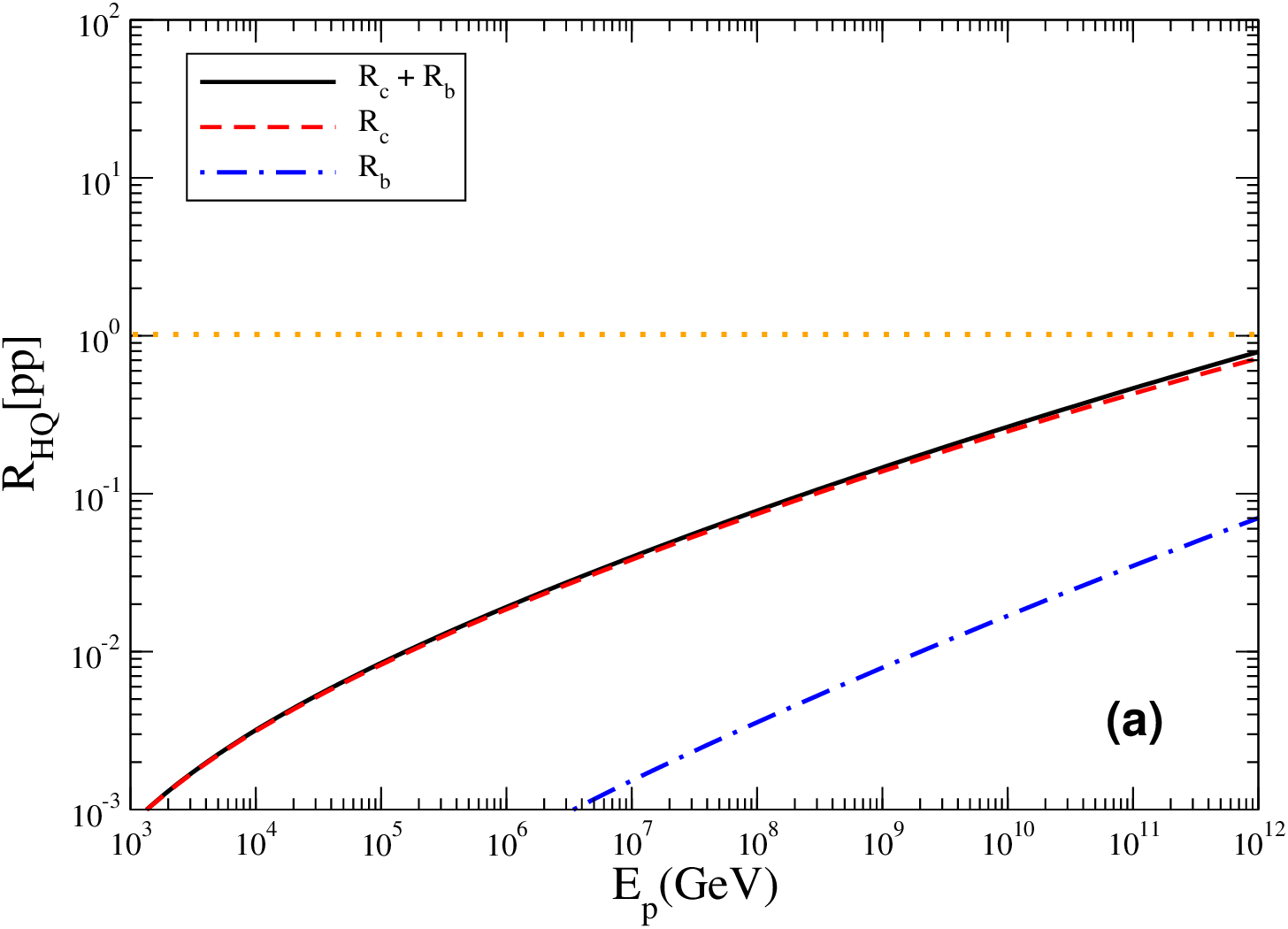}
 \includegraphics[scale=0.25]{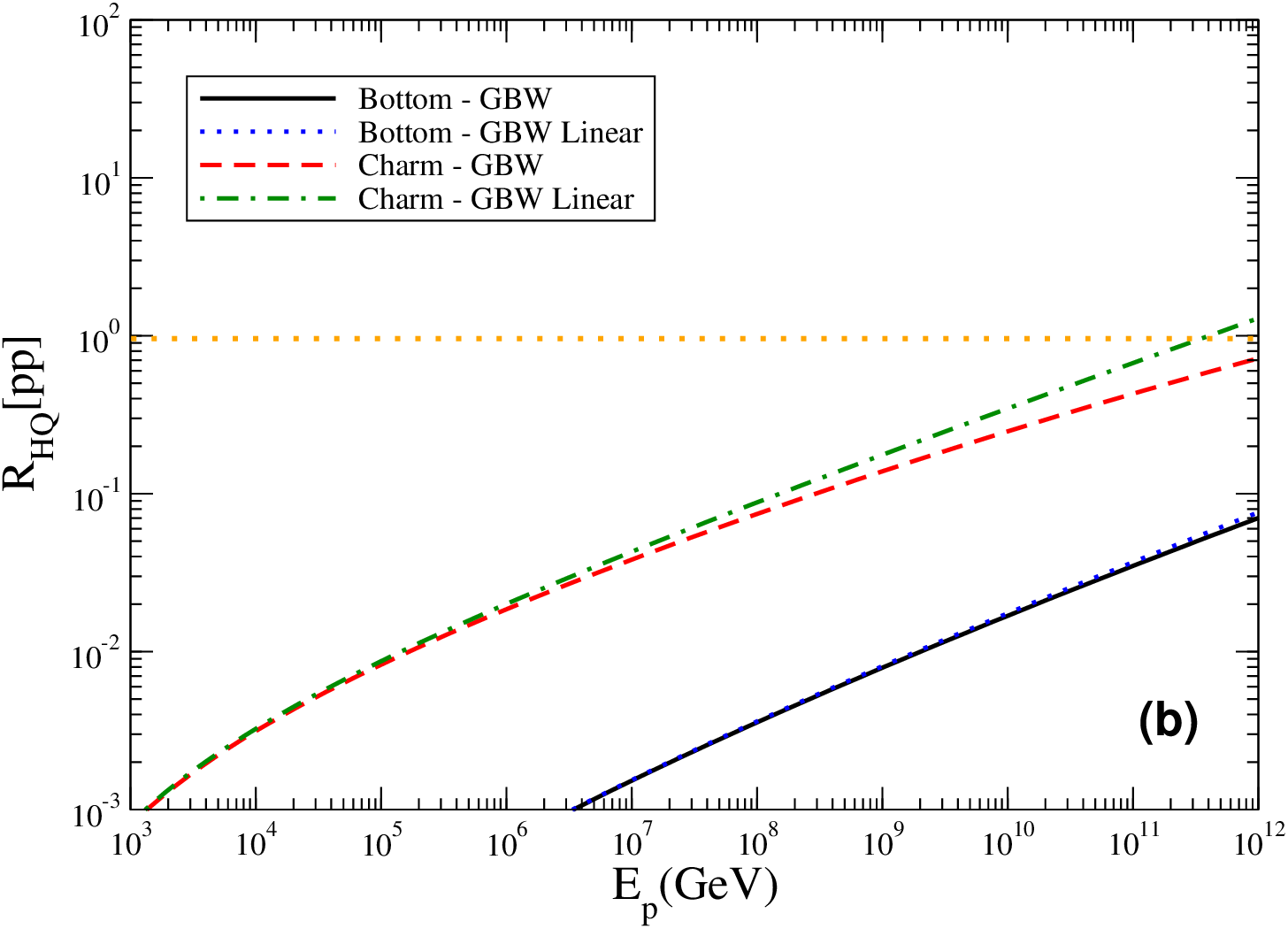}
\caption{(Color online) (a)  The ratio $R_{HQ}[pp]$ as a function of the energy $E_{p}$ of the primary proton. (b) Comparison between the linear and nonlinear predictions for the ratio $R_{HQ}[pp]$. }
\label{fig:protonp}
\end{figure}

\begin{figure}[t]
\vspace{0.5cm}
\includegraphics[scale=0.25] {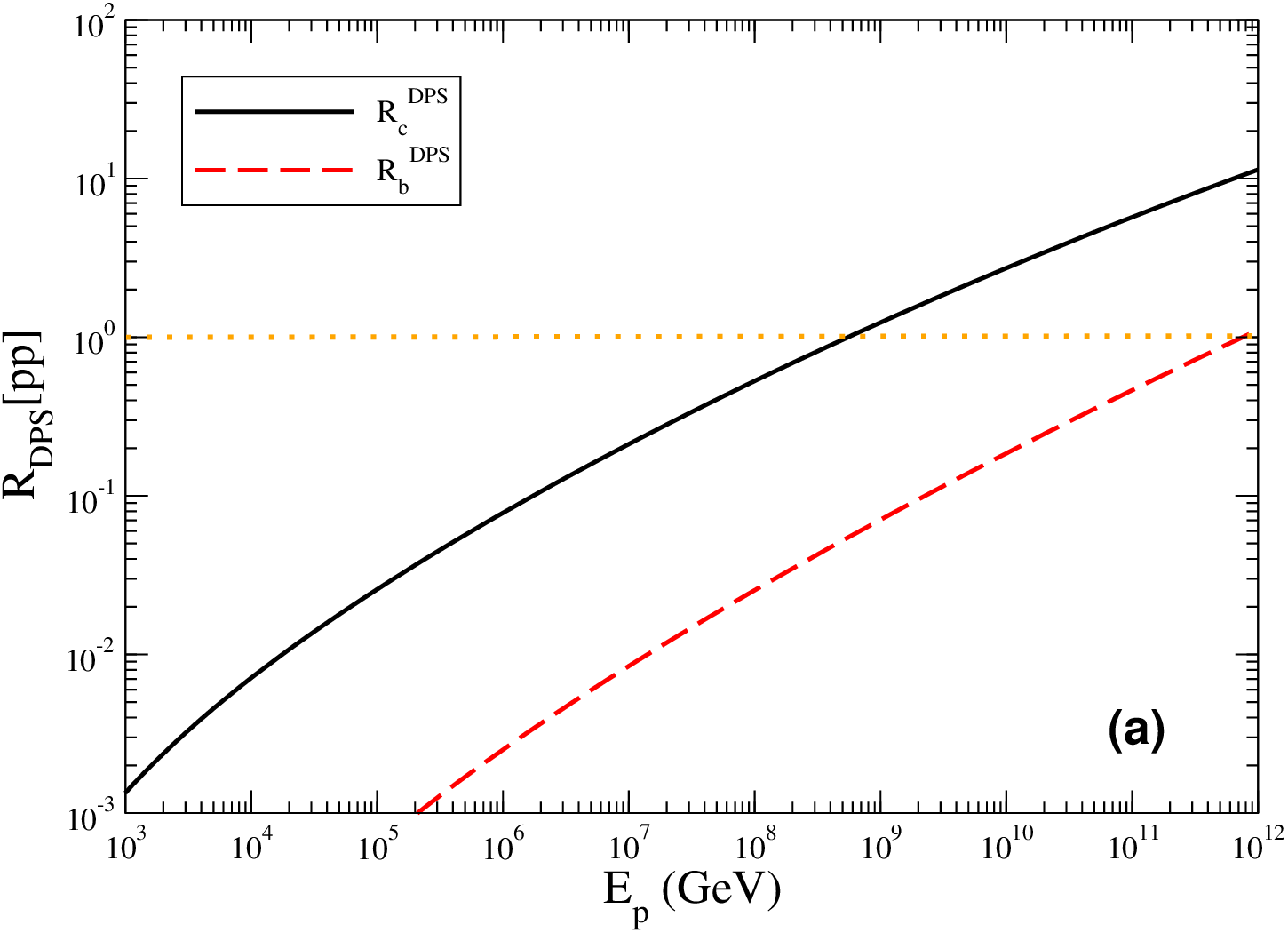}
\includegraphics[scale=0.25]{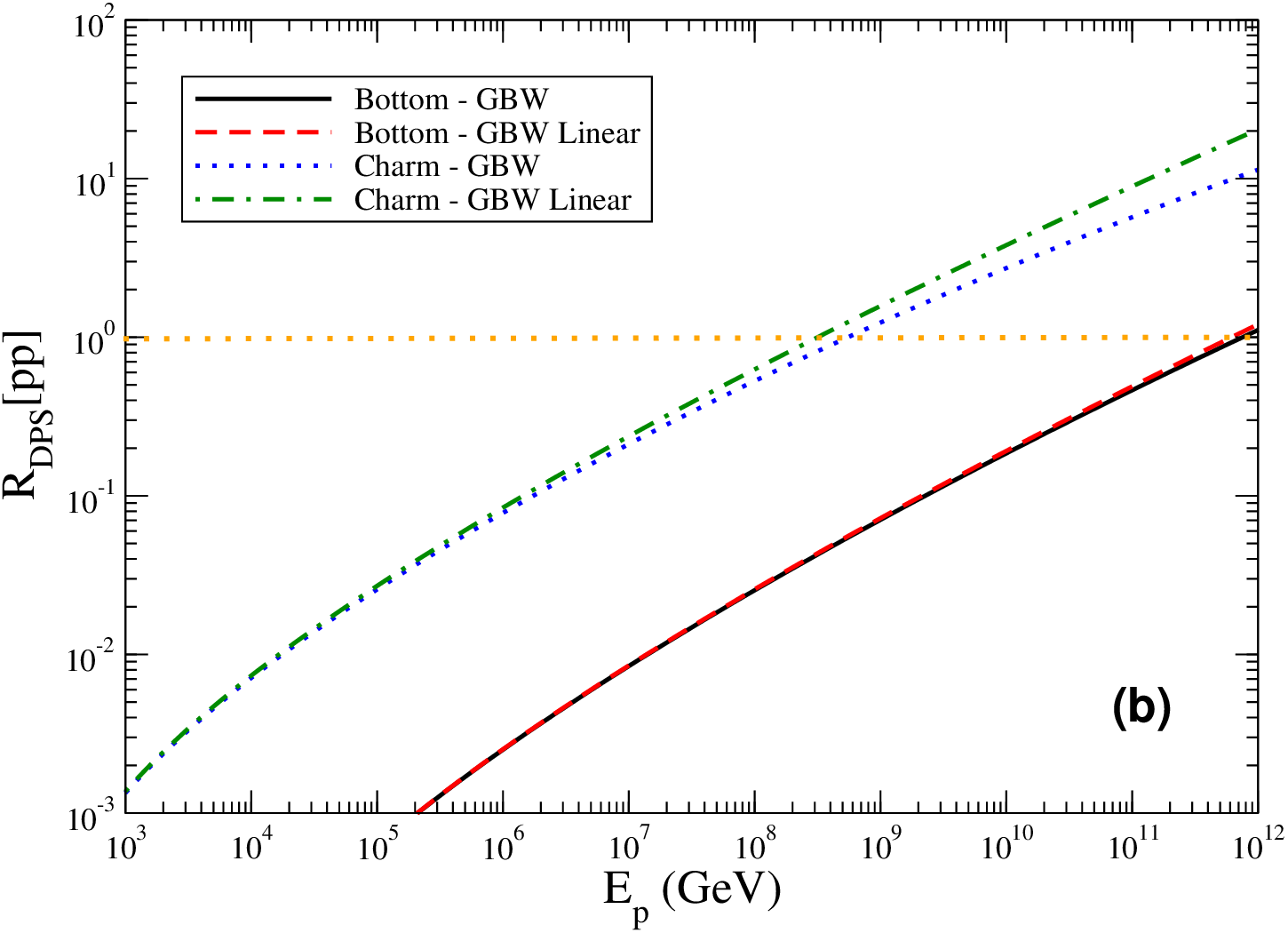}
\caption{(Color online) (a)  The ratio $R_{DPS}[pp]$ { for heavy quark production process} as a function of the energy $E_{p}$ of the primary proton. (b) Comparison between the linear and nonlinear predictions for the ratio $R_{DPS}[pp]$. }
\label{fig:protonp2}
\end{figure}

\section{Results}
\label{results_section}

 In  order to estimate the heavy quark contribution for the total cross sections at ultra high energy cosmic ray interactions, in what follows  we estimate the energy dependence of the ratio
\begin{eqnarray}
R_{HQ}[i p] = \frac{\sigma^{ip}_{HQ}}{\sigma^{ip}_{tot}} \,\,\,
\end{eqnarray}
where $i$ characterizes the primary cosmic ray, which can be either a photon, neutrino or a proton. Moreover, $\sigma^{ip}_{HQ}$ is the total heavy quark  cross section for the production of a given flavor.

Initially let us consider photon - proton interactions ($i = \gamma$). In order to calculate the ratio $R_{HQ}[\gamma p]$ it is necessary to specify the total cross section for the process. Following Ref. \cite{tim} we assume that 
\begin{eqnarray}
\sigma_{tot}^{\gamma p} (W^2) = \sum_{f = u,d,s,c,b,t} \sigma_{\gamma p}^f \,\,\,[\mbox{Eq. (\ref{dipapprox})}]  +  \sigma_{\gamma p}^R (W^2) \label{gamap} 
\end{eqnarray}
where $\sigma_{\gamma p}^R$, associated to the reggeon contribution,  allows to extend the color dipole predictions for the $\gamma p$ cross section down to low values of the photon - proton center-of-mass energy.  As in Ref. \cite{tim} we assume that $\sigma_{\gamma p}^R = A_{\gamma p} .(W^2)^{-\eta}$, with $A_{\gamma p} = 0.135$ mb and $\eta = 0.3$. Moreover,  $m_u = m_d = m_s = 0.21$ GeV, $m_c = 1.5$ GeV, $m_b = 4.5$ GeV and $m_t = 174$ GeV.  In Fig. \ref{fig:gamap} (a) we  demonstrate that this model is able to describe the current experimental data and that high energy behaviour is determined by the dipole contribution, Eq. (\ref{dipapprox}), denoted GBW in the figure. We also present, for the first time, a comparison with the prediction of the COMPETE Collaboration \cite{cudell}, as described in \cite{pdg}. The dipole model predicts a milder energy growth in comparison with the COMPETE prediction, which assumes that $\sigma_{\gamma p} \propto  log^2 W^2$ at very high energies. It is important to emphasize that the calculation of $\sigma_{tot}^{\gamma p}$ constrain all free parameters of the color dipole prediction. As a consequence, our predictions for the heavy quark cross section are parameter free. We have verified that these predictions describe the scarce HERA data for the photoproduction of heavy quarks, as already observed in Ref. \cite{victor_jhep}.
In Fig. \ref{fig:gamap} (b) we present our predictions for $R_{HQ}[\gamma p]$ as a function of the energy $E_{\gamma}$ of the primary photon. We obtain that heavy quark contribution increases with the energy, with the charm contribution ($R_c$) being $\approx$ 20 \% at $E_{\gamma} = 10^{11}$ GeV. On the other hand, the bottom contribution ($R_b$) is of the order of 1 \% and the top contribution is negligible in the full energy range. 

In Fig. \ref{fig:neutrinop} (a) and (b) we present our predictions for the ratio $R_{HQ}[\nu p]$ as a function of the energy $E_{\nu}$ of the primary neutrino for NC and CC interactions, respectively. Our predictions for the total $\nu p$ cross sections have been presented in Refs. \cite{vichepp,diego_victor}, where the model dependence in our predictions have been discussed in detail. However, it is important to emphasize that heavy quark contributions were not explicitly discussed, being presented here for the first time. In the NC case, we predict that the total HQ contribution ($R_c + R_b + R_t$) is $\approx 37$ \% at $E_{\nu} = 10^{11}$ GeV, with the charm and bottom contributions being almost identical and the top one being of the order of 1 \%. On the other hand, for CC interactions, we predict that the total HQ contribution ($R_{cs} + R_{bt}$) is $\approx 52$ \% at $E_{\nu} = 10^{11}$ GeV. 
In particular, for neutrinos with energy of the order of PeV, as those recently probed at IceCube, we predict that the total HQ contributions are $\approx$  25 \% and 44 \% for NC and CC interactions, respectively. The heavy quark contribution for the total $\nu p$ cross section has been estimated previously in Ref. \cite{kms}  considering a unified DGLAP/BFKL approach for the QCD dynamics and the $k_T$ factorization to calculate the structure functions. In Fig.  \ref{fig:neutrinop2} (a) and (b) we present a comparison between our predictions and those presented in Ref. \cite{kms} (denoted KMS hereafter). For NC interactions, we predict smaller/larger values for the charm/bottom production. In contrast, for CC interactions, we predict  very similar values for the $cs$ contribution at ultra high energies and   that the $bt$ contribution is $\approx 75$ \% smaller than those presented in Ref. \cite{kms}. In comparison with the predictions obtained in Ref.  \cite{reno}, where the quark mass effects in the total $\nu p$ cross section were estimated considering the collinear factorization and parton distributions which are solution of the  DGLAP equation, our predictions for high energies are similar.

Let us now consider heavy quark production in proton - proton collisions at ultra high energies. In Refs. \cite{erike_sps,erike_dps}  this approach has been discussed in detail and the heavy quark cross sections at collider energies were estimated. In particular, in Ref. \cite{erike_sps} we have analysed the mass and factorization scale dependence of  the heavy quark cross sections in the color dipole approach and demonstrated that our predictions at LHC energies were strongly dependent on the choice of these parameters. On the other hand, in Ref. \cite{erike_dps} we demonstrated that the description of the  LHC data reduces this uncertainty, with  the GBW model  being able to describe the recent data for the heavy quark production at LHC if we assume $m_c =1.5$ GeV, $m_b = 4.5$ GeV and that the factorization scale is given by $\mu = 2 m_Q$.  As a consequence, we believe that the uncertainty present in our predictions for the heavy quark production at ultra high cosmic ray interactions using the color dipole approach should be small, being mainly associated to the description of the QCD dynamics in this energy regime. 
In Fig. \ref{fig:protonp} (a) we present our predictions for the ratio $R_{HQ}[p p]$ as a function of the energy $E_{p}$ of the primary proton, obtained assuming that  the total $pp$ cross section is given by the parametrization proposed by the COMPETE Collaboration \cite{cudell}, as described in \cite{pdg}. 
Our results demonstrate that the charm contribution for the total cross sections is larger than 10 \% at $E_p \ge 10^8$ GeV  and that at very high energies $\approx 10^{11}$ GeV it becomes of the order of the total $pp$ cross section. 
In order to estimate the magnitude of the gluon saturation effects in the behaviour of the ratio, in Fig. \ref{fig:protonp} (b) we present a comparison between the GBW predictions and those obtained disregarding the nonlinear effects in the dipole - proton scattering amplitude, which implies  ${\cal{N}}(r,Y)=r^2Q_{\mathrm{sat}}^2(Y)/4$ (denoted GBW linear hereafter). For the bottom production, we obtain that these effects can be disregarded. In contrast, for charm production, the  gluon saturation effects diminish the ratio by $\approx 35$ \% for $E_p = 10^{12}$ GeV. In particular, the ratio becomes of order of one at smaller energies if these effects are disregarded. Clearly, these results indicate that new dynamical effects, beyond those included in the dipole factorization  [Eq. (\ref{ccppdip})] and in the QCD dynamics through the gluon saturation effects, should be considered in the kinematical regime probed in UHECR interactions in order to obtain reliable predictions for the charm production. In particular, these results can be interpreted as a signal of the  breakdown of the factorization, which are expected at very high energies as already discussed in the last paragraph of Section \ref{dynamics_section}. Alternatively, if the charm production cross section is measured or determined from the analysis of the flux of prompt leptons at ultra high energies and the value found is of the order of our predictions, then the COMPETE prediction, which is inspired in the Froissart bound,  underestimate the magnitude of the total cross section at cosmic ray energies.

A possible new effect which can contribute for ultra high energies $pp$ interactions and that is not included in our previous analysis, is the heavy quark production in multiple gluon - gluon interactions.  
The  high density of gluons in the initial state of $pp$ collisions at high energies  implies that the probability of  these multiple gluon - gluon interactions  within one proton - proton collision increases. In particular, the probability of having two or more hard interactions in a collision is not significantly 
suppressed with respect to the single interaction probability. Recent theoretical \cite{Marta_Rafal,rafal,erike_dps} and experimental \cite{lhcb_jhep} results demonstrate that the contribution of double parton scattering (DPS) process for the heavy quark production is not negligible already for the LHC energies. Consequently, it is important to estimate the magnitude of this new mechanism for ultra high energy cosmic ray interactions. Following \cite{erike_dps} we  assume that the DPS contribution to  the heavy quark cross section can be expressed in a simple generic form given by
\begin{eqnarray}
\sigma_{p p \rightarrow Q\bar{Q}Q\bar{Q}}^{DPS} = \left( \frac{1}{2} \right) \frac{\sigma^{SPS}_{p p  \rightarrow Q\bar{Q}} 
\sigma^{SPS}_{p p \rightarrow Q\bar{Q}}}{\sigma_{eff}} \,\,,
\label{dps_fac}
\end{eqnarray}
where $Q = c$ or $b$, $\sigma^{SPS}$ is the cross section for the heavy quark production considering the mechanism of single-parton scattering (SPS),  calculated using the Eq. (\ref{ccppdip}), and  $\sigma_{eff}$ is a normalization cross section representing the effective transverse overlap of partonic interactions that produce the DPS process, which we assume as being $\sigma_{eff} = 15$ mb. Moreover, $1/2$ is a 
combinatorial factor which accounts  for  indistinguishable   final states.   The Eq. (\ref{dps_fac}) expresses  the DPS cross section as the product of two individual SPS cross sections assuming that the two SPS sub-processes are uncorrelated and do not interfere (See \cite{erike_dps} for more details).  In Fig. \ref{fig:protonp2} (a) we present our predictions for the  energy dependence of the ratio  $R_{DPS} \equiv \sigma ^{DPS}/\sigma ^{SPS}$. The DPS mechanism for charm production becomes of the order of the SPS one at $E_p = 4 \cdot  10^8$ GeV and dominates at higher energies. For bottom production, the ratio only becomes 1 for $E_p \approx 10^{12}$ GeV.  The magnitude of the gluon saturation effects is estimated in Fig. \ref{fig:protonp2} (b). As expected from our previous results for the heavy quark production in the SPS process, these effects are negligible for the bottom production and diminishes the ratio $R_{DPS}$ by a factor $\approx$ 2 for $E_p = 10^{12}$ GeV. 
Our results indicate that the DPS mechanism cannot be disregarded at cosmic ray energies and should be  included in the 
air-shower simulators in order to obtain reliable predictions for the production of heavy hadrons and consequent flux of prompt leptons at the Earth (For recent studies see Refs. \cite{garcia,bueno}). In  particular, we expect an enhancement of this flux at high energies in comparison with previous studies \cite{leptonflux}, which could have implications at the energies probed at  IceCube. This subject deserves more detailed studies in the future.

\section{Summary}
\label{summary_section}

 Heavy quark production in hard collisions of hadrons, leptons, and photons has been considered as a clean test of
perturbative QCD. This process provides not only many tests of perturbative QCD, but also some of the most important backgrounds to new physics processes,  which have motivated  comprehensive phenomenological studies carried out  at DESY-HERA, Tevatron and LHC. With the advent of the Pierre Auger and  IceCube observatories, the extension of these studies for ultra high energy cosmic ray interactions becomes fundamental. 
In this paper we estimate the heavy quark production at these energies  considering the primary particle as being either a photon, neutrino or a proton. For the first time, the heavy quark contribution for the total cross section  is estimated considering a unique framework -- the dipole formalism --- and taken into account saturation effects, associated to the physical process of parton recombination. Our results indicate that the contribution of heavy quarks is in general non-negligible and can be dominant depending of the process considered. In particular, for $pp$ collisions, the large value obtained for the charm contribution at cosmic ray energies can be interpreted as a signal of the breakdown of the factorization used to calculate the heavy quark cross section.  Moreover, they indicate that new dynamical mechanisms, as for instance, double parton scattering processes, should be included in order to obtain reliable predictions.



\section*{Acknowledgments } 
This work was  partially financed by the Brazilian funding
agencies CNPq, CAPES and FAPERGS.

\end{document}